\documentclass[preprint,apl]{revtex4}
\usepackage{graphicx}
\usepackage{dcolumn}
\usepackage{bm}

\begin{document}

\title{Structures and magnetic properties of ZnO nanoislands}

\author{Yu Yang$^{1,2}$, Ping Zhang$^1$\footnote{Corresponding
author. E-mail address: zhang\underline{ }ping@iapcm.ac.cn}}

\affiliation{$^1$LCP, Institute of Applied Physics and Computational
Mathematics, P.O. Box 8009, Beijing 100088, People's Republic of
China}

\affiliation{$^2$Center for Advanced Study and Department of
Physics, Tsinghua University, Beijing 100084, People's Republic of
China}

\date{\today}

\begin{abstract}
Using first-principles calculations, we systematically study the
atomic structures and electronic properties for two dimensional
triangular ZnO nanoislands that are graphite-like with monolayer and
bilayer thickness. We find that the monolayer ZnO nanoisland with O
terminated zigzag edges is magnetic at its ground state, with the
magnetism coming from the O edge states. The other monolayer and
bilayer ZnO nanoislands with different edge structures are all
nonmagnetic at their ground states. It is further revealed that for
different ZnO nanoislands, their magnetic properties are quite
dependent on their sizes, with larger nanoislands having larger
magnetic moments.\\

\end{abstract}

\maketitle

\clearpage

As a semiconducting metal oxide with a direct band gap, zinc oxide
(ZnO) has vast applications on optoelectronics, transducers and
spintronics \cite{Bagnall1997,Bai2003,Wang2004}. Recently, ZnO
nanostructures of different morphologies have been fabricated and
studied \cite{Kong2003,Gao2005}. Especially, graphite-like hexagonal
ZnO nanofilms have been successfully fabricated \cite{Tusche2007}
and extensively studied for possible usages in electronic devices
\cite{Freeman2006,Goniakowski2007,Pala2007}. Moreover, similar to
the situations in graphene nanostuctures, ZnO nanotubes and
nanoribbons have then also been investigated for possible potential
applications \cite{Wang2007,Shen2007,Mendez2007,Mendez2007NL}. It is
noted that the monolayer ZnO nanoribbons with zigzag edges exhibit
magnetic behavior \cite{Mendez2007NL}. Besides carbon nanotubes and
graphene nanoribbons, two dimensional (2D) graphene nanoislands have
also been paid vast attentions because of their unique flexibility
in tuning magnetic properties \cite{Rossier2007,Wang2008}. Thus when
turn to ZnO nanostructures, one may wonder whether 2D ZnO
nanoislands show tunable magnetic properties or not. Especially when
ZnO nanoisland samples is experimentally attainable by depositing
hexagonal ZnO nanofilms on the Ag(111) surface \cite{Tusche2007}.
Motivated by this question, in this paper we present systematical
investigations on the atomic structures and electronic properties of
2D triangular ZnO nanoislands.

Moreover, the so-called $d^0$-magnetism have recently attracted
extensive interests, which represents for magnetic behaviors in
semiconducting materials (including ZnO) with the absence of
transition metal doping. It was firstly observed in HfO$_2$ thin
films where defects or oxygen vacancies cause magnetic behaviors
\cite{Venkatesan2004}. Later, it has also been theoretically
predicted that carbon doping in ZnO \cite{Pan07} and cation-vacancy
in GaN and BN \cite{Dev08} both will lead to magnetic behaviors. In
addition to these previous researches, here through first-principles
calculations, we further reveal that the monolayer ZnO nanoisland
with O terminated zigzag edge also shows magnetism in its ground
state, even without any vacancies or defects.

In the present work, the first-principles calculations are carried
out using the DMOL$^3$ package. Density functional semicore
pseudopotentials (DSPP) \cite{DSPP} are adopted to replace the
effects of core electrons of Zn and O with a simple potential
including some degree of relativistic correction. The spin-polarized
PW91 \cite{PW91} functional based on generalized gradient
approximation (GGA) is employed to take into account the exchange
and correlation effects of electrons. For valence electrons, the
double-numerical basis with polarized functions is adopted for
expansion of the single-particle wave functions in Kohn-Sham
equations. For better accuracy, the octupole expansion scheme is
adopted for resolving the charge density and Coulombic potential,
and a fine grid is chosen for numerical integration. The charge
density is converged to 1$\times$10$^{-6}$ a.u. in the
selfconsistent calculation. In the optimization process, the energy,
energy gradient, and atomic displacement are converged to
1$\times$10$^{-5}$, 1$\times$10$^{-4}$ and 1$\times$10$^{-3}$ a.u.,
respectively. The atomic charge and magnetic moment are obtained by
the Mulliken population analysis.

In the experiment by Tusche {\it et al.}, the observed triangular 2D
ZnO nanoislands all have the size of about 20 \AA~\cite{Tusche2007}.
So we here start by investigating several kinds of ZnO nanoislands
at this size, including the monolayer and bilayer ZnO nanoislands
with both armchair and zigzag edges. It should be noted that there
are two different kinds of zigzag edges (respectively the Zn- and
O-terminated) for monolayer ZnO nanoislands. In total, five
different ZnO nanoislands are studied here: the monolayer
armchair-edge ZnO nanoisland (AZnONI), and zigzag-edge ZnO
nanoisland with Zn- (Zn-ZZnONI) and O-terminated edges (O-ZZnONI),
the bilayer armchair-edge (BAZnONI) and zigzag-edge ZnO nanoisland
(BZZnONI). The optimized atomic structures for these ZnO nanoislands
are all shown in Fig. 1.

During geometry optimizations, different kinds of ZnO nanoislands
show different structural reconstructions. As shown in Figs.
1(a)-(c), after geometry optimizations for the monolayer ZnO
nanoislands, all the atoms are still inside a same plane. For the
AZnONI, the atoms at the edges and corners are distorted from their
equilibrium places, and nine new Zn-Zn bonds form at the three edges
as shown in Fig. 1(a). For the monolayer zigzag-edge ZnO
nanoislands, the structural reconstructions only occur at the three
corner areas. One can see from Fig. 1(b) that in the Zn-ZZnONI
structure, the two Zn atoms at each corner come close with each
other and form a new Zn-Zn bond, with the three O atoms at the
corners pushed out from their original sites. In the O-ZZnONI
structure shown in Fig. 1(c), the two O atoms at each corner move
away from each other a little making the three Zn atoms at the
corners pulled in from their original sites. The reconstructions for
the BAZnONI and BZZnONI structures are more complex. Figures. 1(e)
and (g) are respectively the side view for the atomic structures in
the blue and red squares in Figs. 1(d) and (f), from which one can
see that the atoms at corners and edges are no longer in the same
plane with the middle atoms. The bond angles at these distorted
areas are decreased to be smaller than 120$^\circ$, indicating that
these atoms incline to transit to the wurtzite configuration.

The electronic properties for the 2D ZnO nanoislands are then
studied, and found to depend critically on their edge structures.
Figure 2 shows the calculated energy levels for the five
geometrically optimized ZnO nanoislands with different edge
structures. From Figs. 2(a), (d), and (e), we can clearly see that
there exist large energy gaps for the AZnONI, BAZnONI and BZZnONI
structures. This is in accordance with the saturation that the ratio
between Zn and O atom numbers is 1:1 in them. The Zn-ZZnONI and
O-ZZnONI structure respectively contains more Zn and O atoms, and
thus are unsaturated. However, spin-splitting effect can only be
observed for the O-ZZnONI structure as shown in Fig. 2(c). And the
calculated total spin ($S_{\rm tot}$) is 2. From the calculated
energy levels shown in Fig. 2(c), we see that the energy difference
between the highest occupied molecular orbital (HOMO) of spin-up and
spin-down electrons ($E_d$ (HOMO)) is only 0.13 eV, which is very
small and indicates that the spin-splitting effect is very weak. The
smaller energy gap in individual gaps of spin-up and -down
electronic states is defined as $E_g^*$, which is 0.18 eV and zero
for the Zn- and O-ZZnONI structure respectively. The zero energy gap
indicates that the O-ZZnONI structure is metallic at its ground
state. And the small energy gap of 0.18 eV indicates that the
Zn-ZZnONI structure is chemically not very stable. However, the Zn-
and O-ZZnONI structures still might be fabricated using some special
methods, since unpolarized hexagonal ZnO monolayers have been
successfully produced already \cite{Tusche2007}.

The deformation electron density and spin density are then further
analyzed to study the origin of the magnetism in the O-ZZnONI
structure. From the deformation electron density shown in Fig. 3(a),
we can see that all the O atoms get some electrons from Zn atoms in
the nanoisland. Mulliken charge density analysis further shows that
the O atoms at the corners and edges get about 0.1 $e$ more
electrons than the O atoms in the middle of the nanoisland. These
extra electrons might supply the excessive electrons in one spin
state. Detailed wavefunction analysis proves that the electronic
states around the Fermi energy are mainly contributed by the
oxygen-dominated edge states. Figure 3(b) shows the spin density
distribution in the O-ZZnONI structure. In this case, it is clear
that the magnetic behavior is due to the oxygen edge states.

The fact that the O-ZZnONI structure without any magnetic impurities
is magnetic might help us better understand the profound origin of
magnetism in ZnO-based nanostructures. Besides, the magnetism in the
O-ZZnONI structure also hints that it might has direct applications
in nanoscale spintronics. So we further investigate the size effects
on the magnetism in different O-ZZnONI structures, which can be
differentiated by the number of O atoms at each edge ($i$), and we
will use O$i$-ZZnONI to represent the O-ZZnONI structure of
different sizes. Based on this definition, the previously discussed
O-ZZnONI structure is actually the O$6$-ZZnONI structure.

Table I shows the calculated magnetic and electronic properties for
some O$i$-ZZnONI structures. We find that the value of all $E_d$
(HOMO) ranges between 0 and 0.2 eV, which is very small and
indicates that the spin-splitting effect is weak for all the
O$i$-ZZnONI structures. From the calculated $S_{\rm tot}$ listed in
Table I, we can see obvious quantum size effects. Except for
O$3$-ZZnONI, the O$i$-ZZnONI structure with a larger size always has
a larger total spin. From careful wavefunction analysis, we find
that in the O$3$-ZZnONI structure, electrons of the middle oxygen
atoms also contribute to its magnetic states. Considering that the
magnetic states in other O$i$-ZZnONI structures are all contributed
by their O edge states, this explains why the O$3$-ZZnONI structure
has a larger $S_{\rm tot}$ than the O$4$-ZZnONI and O$5$-ZZnONI
structures. However, the abnormal large total spin of the
O$3$-ZZnONI structure is because of its too small size. For most
O$i$-ZZnONI structures, larger ones will have larger total spin.

In summary, we have systematically investigated the structures and
magnetic properties for 2D ZnO nanoislands. It is found that the
structural reconstructions are mainly around the three corners and
three edges for triangular ZnO nanoislands. Among the five ZnO
nanoislands with different edge structures, only the O-ZZnONI
structure is magnetic at its ground state. And its spin density
mainly distributes at the edge oxygen atoms. For the O-ZZnONI
structure with different sizes, we reveal that the magnetic and
electronic properties are quite dependent on their sizes. The
O$i$-ZZnONI structure with a larger size always has a larger
magnetic moment, implying their potential usages in spintronics.

This work was supported by the NSFC under grants No. 10604010 and
No. 60776063.

\clearpage

\begin{table}
\caption{The energy difference of HOMO ($E_d$ (HOMO)), the smaller
energy gap in individual gaps of spin-up and spin-down states
($E_g^*$), and the total spin ($S_{\rm tot}$) for the O$i$-ZZnONI
structure with different sizes.}
\begin{tabular*}{9cm}{c@{\hspace{0.8cm}}c@{\hspace{0.8cm}}c@{\hspace{0.8cm}}c}
  \hline
  $i$ & $E_d$ (HOMO) (eV) & $E_g^*$ (eV) & $S_{\rm tot}$ \\
  \hline
  \hline
  3  & 0.01  & 0.19  & 2  \\
  4  & 0.08  & 0.06  & 1  \\
  5  & 0.17  & 0.06  & 1  \\
  6  & 0.13  & 0.00  & 2  \\
  7  & 0.04  & 0.02  & 3  \\
  \hline
\end{tabular*}\label{Stot}
\end{table}

\clearpage

\noindent\textbf{List of captions} \\

\noindent\textbf{Fig.1}~~~ (Color online). Atomic structures from
top view for the AZnONI (a), Zn-ZZnONI (b), O-ZZnONI (c), BAZnONI
(d) and BZZnONI (f). (e) and (g) Atomic structures from side view
for the selected atoms in (d) and (f). In all the structures, red
and grey balls represent O and Zn atoms respectively. \\

\noindent\textbf{Fig.2}~~~ (Color online). Energy levels of AZnONI
(a), Zn-ZZnONI (b), O-ZZnONI (c), BAZnONI (d) and BZZnONI (e). The
left and right columns in each figure correspond to spin-up and
spin-down electronic states, respectively. \\

\noindent\textbf{Fig.3}~~~ (Color online). Contour plot of the
deformation electron density (a) and isosurface plot of the spin
density (b) for the O-ZZnONI. Red and grey balls respectively
represent O and Zn atoms. The isosurface with the value of 0.01
$e$/\AA$^3$~ is shown in blue. \\

\clearpage

\begin{figure}
\includegraphics[width=1.0\textwidth]{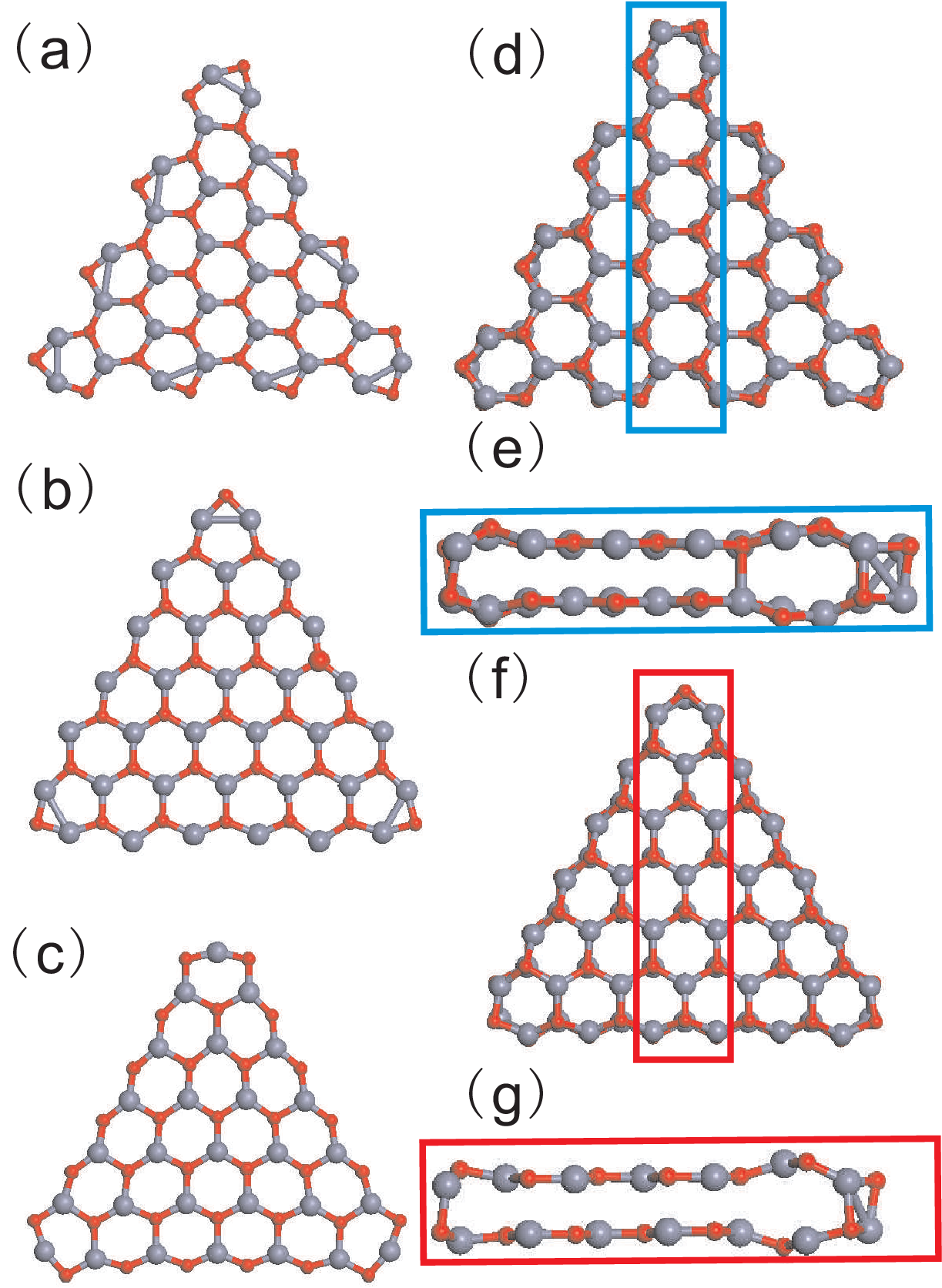}
\caption{\label{fig:fig1}}
\end{figure}
\clearpage
\begin{figure}
\includegraphics[width=1.0\textwidth]{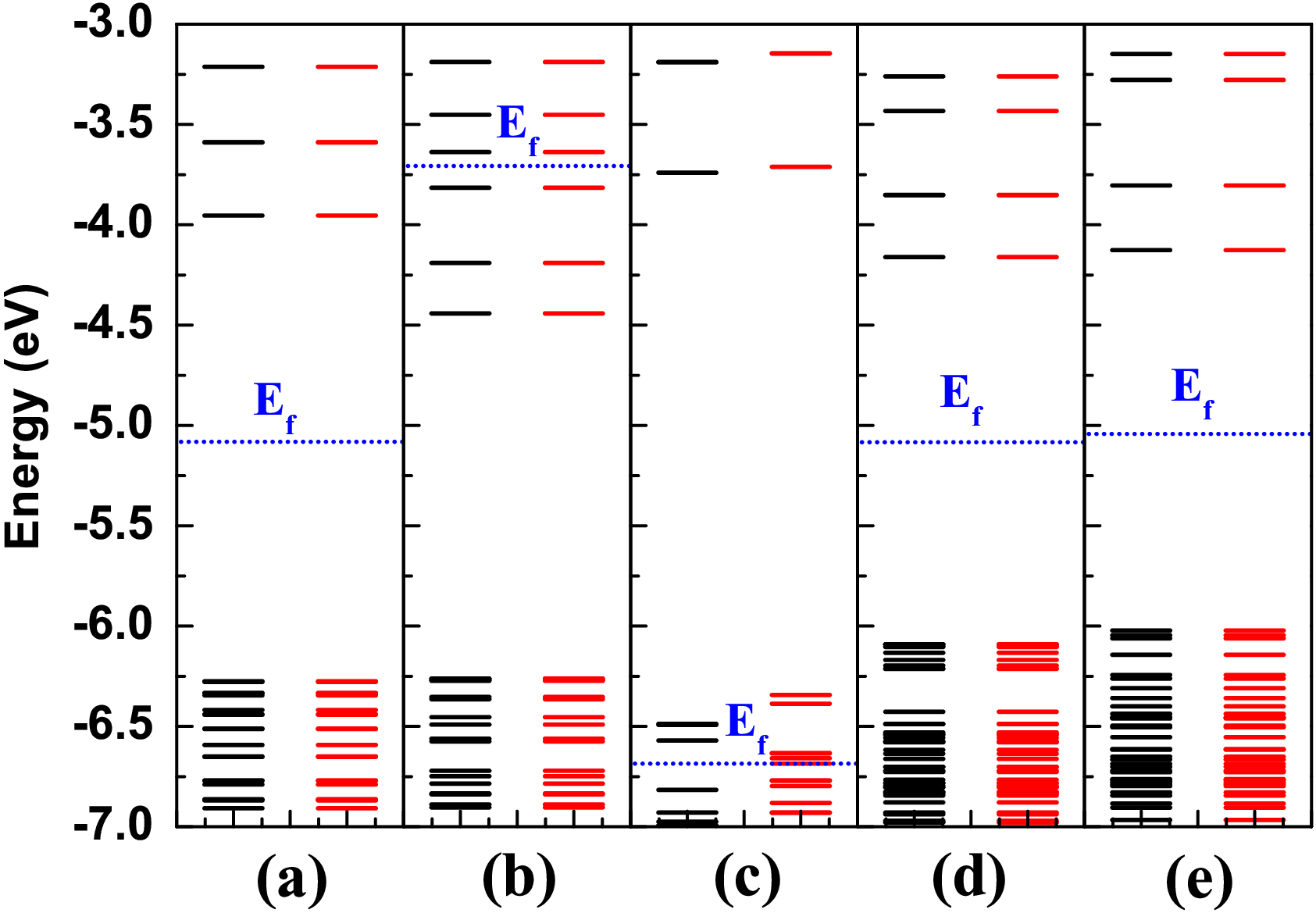}
\caption{\label{fig:fig2}}
\end{figure}
\clearpage
\begin{figure}
\includegraphics[width=1.0\textwidth]{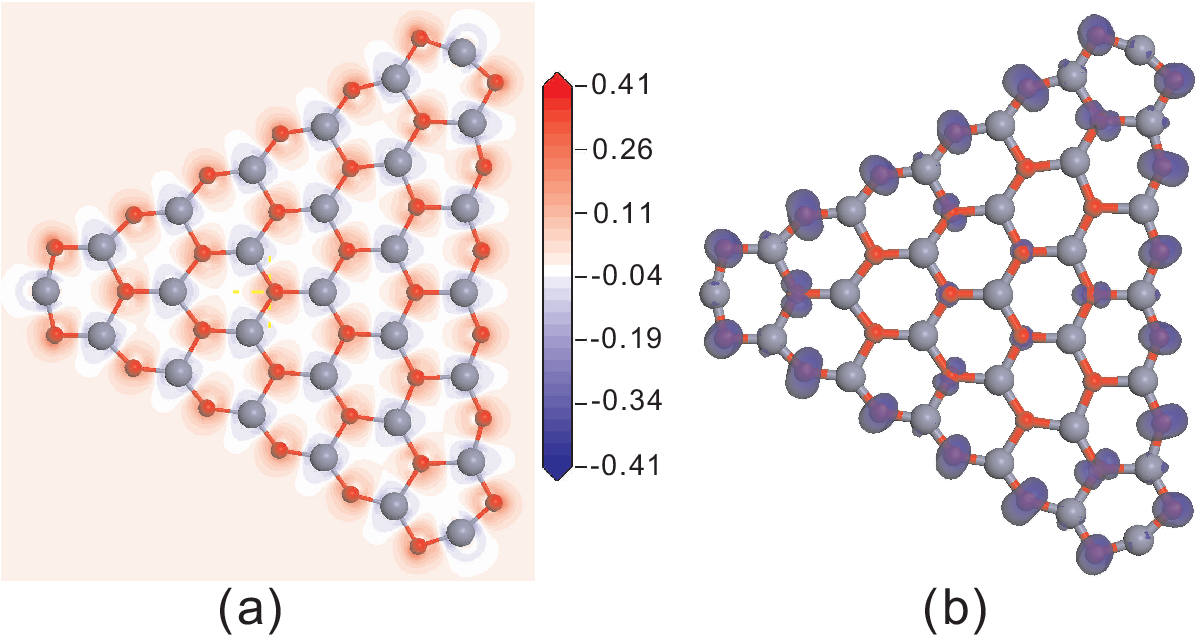}
\caption{\label{fig:fig3}}
\end{figure}
\clearpage
\end{document}